\documentclass[preprint, superscriptaddress, showpacs,preprintnumbers,amsmath,amssymb,prb]{revtex4}
\usepackage{graphicx}

\newcommand{\ve}{\varepsilon}
\begin{document}

\thispagestyle{empty}

\title{Classical limit of the Casimir interaction for thin films
with applications to graphene}

\author{
G.~L.~Klimchitskaya}
\affiliation{Central Astronomical Observatory
at Pulkovo of the Russian Academy of Sciences,
St.Petersburg, 196140, Russia}
\affiliation{Institute of Physics, Nanotechnology and
Telecommunications, St.Petersburg State
Polytechnical University, St.Petersburg, 195251, Russia}
\author{
 V.~M.~Mostepanenko}
\affiliation{Central Astronomical Observatory
at Pulkovo of the Russian Academy of Sciences,
St.Petersburg, 196140, Russia}
\affiliation{Institute of Physics, Nanotechnology and
Telecommunications, St.Petersburg State
Polytechnical University, St.Petersburg, 195251, Russia}

\begin{abstract}
The Casimir interaction between two thin material films, between
a film and a thick plate and between two films deposited on
substrates is considered at large separations (high temperatures)
which correspond to the classical limit. It is shown that the
free energy of the classical Casimir interaction between two
insulating films with no free charge carriers and between an
insulating film and a material plate depends on film thicknesses
and decreases with separation more rapidly than the classical
limit for two thick plates. The free energy of thin films
characterized by the metallic-type dielectric permittivity
decreases as the second power of separation, i.e., demonstrates
the standard classical limit. The obtained results shed
light on the possibility to describe dispersion interaction
between two graphene sheets and between a graphene sheet and
a material plate by modeling graphene as a thin film possessing
some dielectric permittivity. It is argued that the most
reliable results are obtained by describing the reflection
properties on graphene by means of the polarization tensor
in (2+1)-dimensional space-time.
\end{abstract}

\maketitle
\section{Introduction}

It is common knowledge that the van der Waals and Casimir
forces\cite{1,2}
act between closely spaced material surfaces. These forces are
the quantum phenomena caused by the existence of electromagnetic
fluctuations. They are also known under the generic name
{\it dispersion forces}. At the shortest separations between
the test bodies below a few nanometers, where the relativistic
retardation does not play any role, the term
{\it van der Waals forces} is used. At larger separations,
where the retardation effects contribute essentially,
dispersion forces are usually referred to as the
{\it Casimir forces}. During the last few years the Casimir
forces attract much attention as a multidisciplinary subject
having prospective applications in condensed matter physics,
atomic physics, quantum field theory and in astrophysics and
cosmology (see monographs\cite{2,3,4,5}).  Measurements of the
Casimir forces between metallic and semiconductor surfaces
have also attracted considerable interest (see
reviews\cite{6,7,8}).

The fundamental theory of the  van der Waals and Casimir
forces was elaborated by Lifshitz\cite{9} and is called the
{\it Lifshitz theory}. This theory expresses the free energy
and force acting between two thick parallel plates
(semispaces) in thermal equilibrium with an environment at
temperature $T$ in terms of their frequency-dependent
dielectric permittivities. The Lifshitz theory was
generalized\cite{10,11}
for an arbitrary number of plane parallel layers of
magnetodielectrics characterized by the frequency-dependent
dielectric permittivities and magnetic permeabilities.
In the last few years the Lifshitz-type theory was
developed\cite{12,13} which describes the van der Waals and
Casimir interaction between bodies of arbitrary shape.

Considerable recent attention has been focused on physics
of one-atom-thick graphene sheets and other carbon
nanostructures which possess unusual electrical, mechanical
and optical properties of high promise for many
applications.\cite{14,15}
Among different processes and effects which have been
investigated, a lot of papers was devoted to calculation
of dispersion forces between two carbon nanostructures and
between a carbon nanostructure and a body made of some
usual
material.\cite{16,17,18,19,20,21,22,23,24,25,26,27,28,29,30,31}
Many of them used in computations the Lifshitz theory
combined with one or other model for the dielectric permittivity
of graphene.\cite{24,25,26,27,28,29}
Otherwise the reflection coefficients of the electromagnetic
oscillations on graphene were expressed\cite{21,23,30,31} in
terms of the polarization tensor in $(2+1)$-dimensional
space-time in the framework of the Dirac model (an alternative
formalism using Coulomb coupling between density fluctuations
which leads to the same results was also proposed\cite{22}).
In Ref.~\cite{31} the computational results for the Casimir
interaction between two graphene sheets computed using
different models of dielectric permittivity on the one hand and
the polarization tensor on the other hand have been compared
and some disagreements were found. It was concluded that the
origin of these disagreements invites further investigation.

Keeping in mind that when using the concept of dielectric
permittivity graphene is likened to an ordinary material film
of some definite thickness $d$, in this paper we investigate
the thermal Casimir interaction in the presence of thin films.
This problem was already discussed in the literature in relation
to measurements of the Casimir force. Thus, the van der Waals and
Casimir forces acting between Al test bodies covered with Au
films were computed at zero temperature.\cite{32}
In a similar way the role of thin metallic and semiconductor
films at zero temperature was investigated in
Refs.~\cite{33,34,35}.
At room temperature all these results are applicable for films
made of usual materials at separation distances below 1--2
micrometers. They are not applicable to graphene films
possessing large thermal effects even at very short separation
distances.\cite{22} The Casimir force between atomically thin
Au films was also calculated using anisotropic dielectric
functions obtained with the help of density functional
theory.\cite{36} This is in line with some dielectric
functions used for the description of graphene.\cite{26,27}

Below we calculate the Casimir interaction in the presence
of thin films in the so-called {\it classiacl limit}.\cite{37}
This corresponds to the case of large separations (high
temperatures) when only the zero-frequency terms of the Lifshitz
formulas provide the total Casimir free energy and force.
The obtained results are presented in the closed analytic form
and do not depend on a specific choice of the dielectric
permittivity, but rather on its behavior at low frequencies.
This helps to reveal the physical role of film thickness
when applying the concept of the dielectric permittivity
to graphene and simplifies comparison with the calculation
results obtained using the polarization tensor.
Below we consider the thermal Casimir interaction between both
isolated films and films deposited on a substrate.
The dielectric functions covered by our analysis are both of
dielectric and metallic type, i.e., are adapted for theoretical
description of both insulators and metals.

The paper is organized as follows. In Sec.~II we consider the
Casimir interaction between two thin dielectric films and
between a dielectric film and a material semispace (either dielectric
or metallic). In Sec.~III the case of thin metallic films is
investigated. Section~IV is devoted to the Casimir interaction
between two thick plates (semispaces) coated with thin films.
In Sec.~V the comparison of our results for thin films with
respective results for graphene sheets, obtained using the
polarization tensor, are presented. Our conclusions and
discussion are contained in Sec.~VI.

\section{Dielectric film interacting with another dielectric film or
a material plate}

\begin{figure}[b]
\vspace*{-8cm}
\centerline{\hspace*{2cm}
\includegraphics{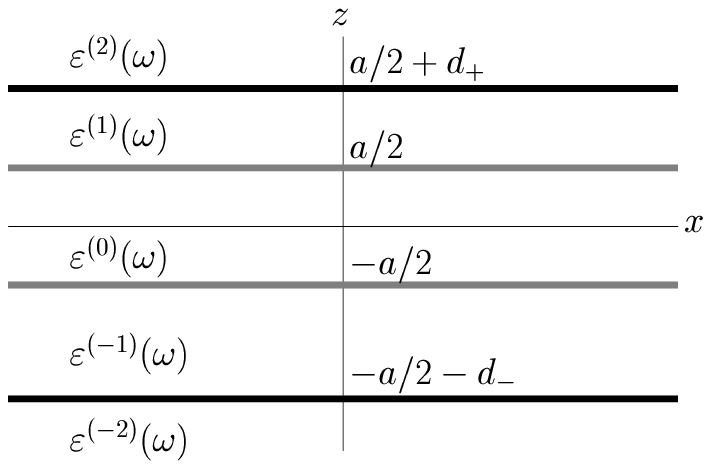}
}
\vspace*{-13cm}
\caption{\label{fg1}
Stratified medium consisting of three layers of finite
thickness $d_{-},\>a$, and $d_{+}$ with dielectric permittivities
$\varepsilon^{(-1)}(\omega)$, $\varepsilon^{(0)}(\omega)$, and
$\varepsilon^{(1)}(\omega)$, enclosed
between two semispaces $z\leq -a/2-d_{-}$ and $z\geq a/2+d_{+}$
with dielectric permittivities
$\varepsilon^{(-2)}(\omega)$ and $\varepsilon^{(2)}(\omega)$,
respectively.
}
\end{figure}

For the needs of this and following sections, we consider the
Casimir interaction between two films of thicknesses $d_{\pm}$
with dielectric permittivities $\ve^{\,(\pm 1)}(\omega)$
separated with a vacuum gap of thickness $a$ with dielectric
permittivity $\ve^{\,(0)}(\omega)=1$. The films are deposited
on two thick plates (semispaces) with respective
dielectric permittivities $\ve^{\,(\pm 2)}(\omega)$
as shown in Fig.~\ref{fg1}. The classical limit\cite{37}
holds at high temperatures or, equivalently, at large separations
satisfying the inequalities
\begin{equation}
T\gg T_{\rm eff}\equiv\frac{\hbar c}{2ak_B}, \qquad
a\gg a_{ T}\equiv\frac{\hbar c}{2k_BT},
\label{eq1}
\end{equation}
\noindent
where $k_B$ is the Boltzmann constant. In fact in these
inequalities
much larger can be replaced with larger (note that at
$T=300\,$K $a_T\approx 3.8\,\mu$m and the classical limit
starts from $a\approx 5\,\mu$m).

In the classical limit we can restrict our consideration to the
zero-frequency term of the Lifshits formula.\cite{2}
Then the Casimir free energy per unit area of the films is
given by\cite{2,10,11,32}
\begin{eqnarray}
&&
{\cal F}(a,T)=\frac{k_BT}{4\pi}\int_{0}^{\infty}\!\!
k_{\bot}dk_{\bot}\sum\limits_{\alpha}
\label{eq2} \\
&&~~~\times
\ln\left[1-R_{\alpha}^{(+)}(0,k_{\bot})
R_{\alpha}^{(-)}(0,k_{\bot})e^{-2k_{\bot}a}\right],
\nonumber
\end{eqnarray}
\noindent
where $k_{\bot}=|\mbox{\boldmath$k$}_{\bot}|$,
$\mbox{\boldmath$k$}_{\bot}$ is the projection of the
wave vector on the plane of films, and
$\alpha={\rm TM,\,TE}$ labels the two independent
polarizations of the electromagnetic field,
transverse magnetic and transverse electric,
respectively.
The reflection coefficients on the planes
$z=\pm a/2$ are given by
\begin{equation}
R_{\alpha}^{(\pm)}(0,k_{\bot})=
\frac{r_{\alpha}^{(0,\pm 1)}(0,k_{\bot})+
r_{\alpha}^{(\pm 1,\pm 2)}(0,k_{\bot})
e^{-2k^{(\pm 1)}(0,k_{\bot})d_{\pm}}}{1+
r_{\alpha}^{(0,\pm 1)}(0,k_{\bot})
r_{\alpha}^{(\pm 1,\pm 2)}(0,k_{\bot})
e^{-2k^{(\pm 1)}(0,k_{\bot})d_{\pm}}}.
\label{eq3}
\end{equation}
\noindent
Here, the reflection coefficients on the various boundary
planes are given by
\begin{eqnarray}
&&
r_{\rm TM}^{(n,n^{\prime})}(i\xi,k_{\bot})=
\frac{\ve^{(n^{\prime})}(i\xi)k^{(n)}(i\xi,k_{\bot})-
\ve^{(n)}(i\xi)k^{(n^{\prime})}(i\xi,k_{\bot})}{\ve^{(n^{\prime})}(i\xi)
k^{(n)}(i\xi,k_{\bot})+\ve^{(n)}(i\xi)k^{(n^{\prime})}(i\xi,k_{\bot})},
\nonumber \\
&&
r_{\rm TE}^{(n,n^{\prime})}(i\xi,k_{\bot})=
\frac{k^{(n)}(i\xi,k_{\bot})-
k^{(n^{\prime})}(i\xi,k_{\bot})}{k^{(n)}(i\xi,k_{\bot})
+k^{(n^{\prime})}(i\xi,k_{\bot})},
\label{eq4}
\end{eqnarray}
\noindent
where $n,n^{\prime}=0,\,\pm 1,\,\pm 2$, $\xi$ is the frequency
along the imaginary frequency axis and
\begin{equation}
k^{(n)}(i\xi,k_{\bot})=\left[k_{\bot}^2+\ve^{(n)}(i\xi)
\frac{\xi^2}{c^2}\right]^{1/2}.
\label{eq5}
\end{equation}
\noindent
Below we also use the dimensionless variable $y=2k_{\bot}a$
instead of $k_{\bot}$.

We begin from the consideration of the Casimir interaction
between two parallel dielectric films in vacuum.
In this case $\ve^{(\pm 2)}(\omega)=1$,
$k^{(n)}(0,k_{\bot})=k_{\bot}$ and from Eq.~(\ref{eq4})
using Eq.~(\ref{eq5}) one obtains
$r_{\rm TE}^{(n,n^{\prime})}(0,k_{\bot})=0$ and following
nonzero TM reflection coefficients
\begin{equation}
r_{\rm TM}^{(0,\pm 1)}(0,k_{\bot})=
-r_{\rm TM}^{(\pm 1,\pm 2)}(0,k_{\bot}) =
\frac{\ve_0^{(\pm 1)}-1}{\ve_0^{(\pm 1)}+1}
\equiv r_0^{(\pm 1)},
\label{eq6}
\end{equation}
\noindent
where $\ve_0^{(n)}\equiv \ve^{(n)}(0)$.
Then for the reflection coefficients (\ref{eq3}) we
arrive at
\begin{eqnarray}
&&
R_{\rm TM}^{(\pm)}(0,k_{\bot})=r_0^{(\pm 1)}
\frac{1-e^{-2k_{\bot}d_{\pm}}}{1-
{r_0^{(\pm 1)}}^2e^{-2k_{\bot}d_{\pm}}},
\nonumber \\
&&
R_{\rm TE}^{(\pm)}(0,k_{\bot})=0.
\label{eq7}
\end{eqnarray}
\noindent
In all subsequent calculations we assume that our films are
thin as compared to separation distance, i.e., $d_{\pm}\ll a$.

Now we rewrite the integral in Eq.~(\ref{eq2}) in terms of
the variable $y$ and expand the quantities
$R_{\rm TM}^{(\pm)}(0,y)$ in powers of a small parameter
$d_{\pm}/a$ preserving only the main contribution
\begin{equation}
R_{\rm TM}^{(\pm)}(0,y)=r_0^{(\pm 1)}
\frac{y}{1-{r_0^{(\pm 1)}}^2}
\frac{d_{\pm}}{a}+O\left(\frac{d_{\pm}^2}{a^2}\right).
\label{eq8}
\end{equation}
\noindent
Substituting Eq.~(\ref{eq8}) in Eq.~(\ref{eq2}),
one obtains
\begin{equation}
{\cal F}(a,T)\approx\frac{k_BT}{16\pi a^2}\int_{0}^{\infty}
\!\!\!ydy\ln\left[1-
\frac{r_0^{(1)}r_0^{(-1)}d_{+}d_{-}}{(1-{r_0^{(1)}}^2)
(1-{r_0^{(-1)}}^2)a^2}y^2e^{-y}\right].
\label{eq9}
\end{equation}
\noindent
Expanding the logarithm in Eq.~(\ref{eq9}) in powers of a small
parameter $d_{+}d_{-}/a^2$ and integrating, we finally
obtain
\begin{equation}
{\cal F}(a,T)\approx-\frac{3r_0^{(1)}r_0^{(-1)}}{8\pi(1-{r_0^{(1)}}^2)
(1-{r_0^{(-1)}}^2)}\,\frac{k_BTd_{+}d_{-}}{a^4}
\label{eq10}
\end{equation}
\noindent
or, for two similar films with $d_{+}=d_{-}=d$ and
$r_0^{(1)}=r_0^{(-1)}\equiv r_0$
\begin{equation}
{\cal F}(a,T)\approx-\frac{3r_0^2}{8\pi(1-r_0^2)^2}
\,\frac{k_BTd^2}{a^4}.
\label{eq11}
\end{equation}
\noindent
This result demonstrates the classical limit, but
an unusually rapid decrease with the increase of separation
(the inverse fourth power instead of the inverse second) if to
compare with the classical limit for the
case of two dielectric semispaces. The physical reason for
this difference is
explained by the presence of two additional dimensional
parameters $d_{\pm}$ and will be further discussed in
Sec.~V in connection with the case of graphene sheets.
{}From Eq.~(\ref{eq10}) for the Casimir pressure one finds
\begin{equation}
P(a,T)\approx-\frac{3r_0^{(1)}r_0^{(-1)}}{2\pi(1-{r_0^{(1)}}^2)
(1-{r_0^{(-1)}}^2)}\,\frac{k_BTd_{+}d_{-}}{a^5}.
\label{eq12}
\end{equation}

Now we consider the Casimir interaction of a thin dielectric
film with a dielectric semispace. In this case
$\ve^{(2)}(\omega)=1$, $\ve^{(-2)}(\omega)=\ve^{(-1)}(\omega)$
and, again, $k^{(n)}(0,k_{\bot})=k_{\bot}$. Thus, from
Eqs.~(\ref{eq3}) and (\ref{eq4}),
$r_{\rm TE}^{(n,n^{\prime})}(0,k_{\bot})=
R_{\rm TE}^{(\pm)}(0,k_{\bot})=0$. For the TM reflection
coefficients we obtain
\begin{eqnarray}
&&
r_{\rm TM}^{(0,1)}(0,k_{\bot})=r_0^{(1)},\quad
r_{\rm TM}^{(1,2)}(0,k_{\bot})=-r_0^{(1)},
\nonumber \\
&&
r_{\rm TM}^{(0,-1)}(0,k_{\bot})=
\frac{\ve_0^{(-2)}-1}{\ve_0^{(-2)}+1}\equiv r_0^{(-2)},
\nonumber \\
&&
r_{\rm TM}^{(-1,-2)}(0,k_{\bot})=0,
\quad
R_{\rm TM}^{(-)}(0,k_{\bot})=r_0^{(-2)},
\nonumber \\
&&
R_{\rm TM}^{(+)}(0,k_{\bot})=r_0^{(1)}
\frac{1-e^{-2k_{\bot}d_{+}}}{1-
{r_0^{(1)}}^2e^{-2k_{\bot}d_{+}}}.
\label{eq13}
\end{eqnarray}
\noindent
Expanding, as above, in powers of a small parameter $d_{+}/a$
and using the variable $y$, we obtain from Eq.~(\ref{eq2})
\begin{equation}
{\cal F}(a,T)\approx\frac{k_BT}{16\pi a^2}\int_{0}^{\infty}
\!\!\!ydy\ln\left[1-
\frac{r_0^{(1)}r_0^{(-2)}d_{+}}{(1-{r_0^{(1)}}^2)a
}ye^{-y}\right].
\label{eq14}
\end{equation}
\noindent
After expansion of the logarithm in powers of the same
parameter one arrives at
\begin{eqnarray}
&&
{\cal F}(a,T)\approx-\frac{r_0^{(1)}r_0^{(-2)}}{8\pi(1-{r_0^{(1)}}^2)}
\,\frac{k_BTd_{+}}{a^3},
\nonumber \\
&&
P(a,T)\approx-\frac{3r_0^{(1)}r_0^{(-2)}}{8\pi(1-{r_0^{(1)}}^2)}
\,\frac{k_BTd_{+}}{a^4}.
\label{eq15}
\end{eqnarray}
\noindent
These results again demonstrate the classical limit with some
specific dependence on separation caused by the fact that one
of the semispaces was replaced with a thin film (the comparison
with the case of a graphene sheet interacting with a dielectric
plate is contained in Sec.~V).

Next we deal with the classical Casimir interaction between the
same dielectric film and a metallic semispace. In this case,
again, $\ve^{(2)}(\omega)=1$, $\ve^{(-2)}(\omega)=\ve^{(-1)}(\omega)$
and $k^{(1)}(0,k_{\bot})=k^{(2)}(0,k_{\bot})=k_{\bot}$. Thus, from
Eqs.~(\ref{eq3}) and (\ref{eq4}), we have
$r_{\rm TE}^{(n,n^{\prime})}(0,k_{\bot})=
R_{\rm TE}^{(+)}(0,k_{\bot})=0$, where $n$ and $n^{\prime}$ here
take the values 0,\,1,\,2 and also $n=-1$, $n^{\prime}=-2$.
Note that the value of the coefficient
$r_{\rm TE}^{(0,-1)}(0,k_{\bot})=r_{\rm TE}^{(0,-2)}(0,k_{\bot})$
depends on the used model of metal (see below) and can be not
equal to zero leading to a nonzero value of
$R_{\rm TE}^{(-)}(0,k_{\bot})$. This, however, does not influence
on the result due to Eq.~(\ref{eq2}). For the TM reflection
coefficient $R_{\rm TM}^{(+)}(0,k_{\bot})$ we obtain the result
already presented in Eq.~(\ref{eq13}).
As to $R_{\rm TM}^{(-)}(0,k_{\bot})$, it is easily calculated
taking into account that for any model of the dielectric permittivity
of a metal $|\ve^{(-2)}(\omega)|\to\infty$ when $\omega\to 0$
leading to
\begin{equation}
r_{\rm TM}^{(0,-1)}(0,k_{\bot})=R_{\rm TM}^{(-)}(0,k_{\bot})=1.
\label{eq16}
\end{equation}
\noindent
Thus, the Casimir free energy per unit area after an expansion
in small parameter $d_{+}/a$ takes the form
\begin{equation}
{\cal F}(a,T)\approx\frac{k_BT}{16\pi a^2}\int_{0}^{\infty}
\!\!\!ydy\ln\left[1-
\frac{r_0^{(1)}d_{+}}{(1-{r_0^{(1)}}^2)a
}ye^{-y}\right]
\label{eq17}
\end{equation}
\noindent
resulting in
\begin{eqnarray}
&&
{\cal F}(a,T)\approx-\frac{r_0^{(1)}}{8\pi(1-{r_0^{(1)}}^2)}
\,\frac{k_BTd_{+}}{a^3},
\nonumber \\
&&
P(a,T)\approx-\frac{3r_0^{(1)}}{8\pi(1-{r_0^{(1)}}^2)}
\,\frac{k_BTd_{+}}{a^4}.
\label{eq18}
\end{eqnarray}
\noindent
This can be also obtained from Eq.~(\ref{eq15}) in the limiting
case $r_0^{(-2)}\to 1$, as it should be when the dielectric plate
is replaced by a metallic.

\section{Interaction between two metallic films}

We now turn our attention to the case of two parallel metallic
films. As discussed above, in this case
$|\ve^{(\pm 1)}(\omega)|\to\infty$ when $\omega\to 0$.
The low-frequency behavior of metals is commonly described
by the Drude dielectric function
\begin{equation}
\ve_D^{(\pm 1)}(\omega)=1-
\frac{\omega_{p,\pm 1}^2}{\omega[\omega+i\gamma_{\pm 1}(T)]},
\label{eq19}
\end{equation}
\noindent
where $\omega_{p,\pm 1}$ are the plasma frequencies and
$\gamma_{\pm 1}(T)$ are the relaxation parameters for the upper
and lower films, respectively. At sufficiently low frequencies
it holds
$\omega\ll\gamma_{\pm 1}(T)$ and Eq.~(\ref{eq19}) results in
$\ve_D^{(\pm 1)}(\omega)\sim\omega^{-1}$, as it should be
at quasistatic $\omega$ in accordance to the Maxwell
equations. It was demonstrated, however, in several experiments
with metallic test bodies\cite{38,39,40,41,42} that theoretical
predictions of the Lifshitz theory using Eq.~(\ref{eq19}) are
excluded by the measurement data. The data were found in
excellent agreement with the low-frequency behavior given by
the plasma model
\begin{equation}
\ve_p^{(\pm 1)}(\omega)=1-
\frac{\omega_{p,\pm 1}^2}{\omega^2}.
\label{eq20}
\end{equation}
\noindent
This model is in fact justified in the frequency region of
infrared optics $\omega\gg\gamma_{\pm 1}(T)$ where the
relaxation properties do not play any role. What is more,
the Lifshitz theory using Eq.~(\ref{eq19}) was
shown\cite{43,44,45} to violate the Nernst heat theorem
whereas the same theory is thermodynamically consistent when
using Eq.~(\ref{eq20}). (We note that in two other
experiments\cite{46,47} the Lifshitz theory using
Eq.~(\ref{eq19}) was confirmed, but these experiments are not
direct measurements of the Casimir force and their results were
disputed in the literature.\cite{48,49,50,51,52})
Keeping in mind that the fundamental reasons behind preference
of the dielectric permittivity (\ref{eq20}) in the equilibrium
fluctuational electrodynamics remain unclear, below we perform
all calculations using both Eqs.~(\ref{eq19}) and (\ref{eq20})
when the result is sensitive to the low-frequency behavior
of $\ve$.

First, we consider the two films described at low frequencies
by the Drude model (\ref{eq19}). In this case it holds
\begin{equation}
\lim_{\xi\to 0}\ve^{(n)}(i\xi)\xi^2=0
\label{eq21}
\end{equation}
\noindent
and $k^{(n)}(0,k_{\bot})=k_{\bot}$. As a result, from
Eqs.~(\ref{eq3}) and (\ref{eq4}) one obtains
\begin{equation}
r_{\rm TE}^{(n,n^{\prime})}(0,k_{\bot})=
R_{\rm TE}^{(\pm)}(0,k_{\bot})=0
\label{eq22}
\end{equation}
\noindent
and for the TM mode
\begin{equation}
r_{\rm TM}^{(0,\pm 1)}(0,k_{\bot})=
-r_{\rm TM}^{(\pm 1,\pm 2)}(0,k_{\bot})=
R_{\rm TM}^{(\pm)}(0,k_{\bot})=1.
\label{eq23}
\end{equation}
\noindent
Substituting Eqs.~(\ref{eq22}) and (\ref{eq23}) in
Eq.~(\ref{eq2}) written in terms of the variable $y$, we find
the Casimir free energy per unit area calculated using the
Drude model
\begin{eqnarray}
&&
{\cal F}_D(a,T)=\frac{k_BT}{16\pi a^2}\int_{0}^{\infty}\!\!\!
ydy\ln(1-e^{-y})
\nonumber \\
&&
\phantom{{\cal F}_D(a,T)}
=-\frac{k_BT\zeta(3)}{16\pi a^2},
\label{eq24}
\end{eqnarray}
\noindent
where $\zeta(z)$ is the Riemann zeta function.
Note that the result (\ref{eq24}) does not depend on the film
thicknesses and coincides with respective result for two
metallic semispaces described by the Drude model.\cite{2}
In the high-temperature limit the same result was also
obtained\cite{36} for atomically thin Au films described
by the anisotropic dielectric functions.

Now we consider the two films described at low frequencies
by the plasma dielectric function. This case is more
interesting and presents several possibilities depending on
the values of involved parameters. Concerning the TM mode,
the result in Eq.~(\ref{eq23}) and, thus, in Eq.~(\ref{eq24})
remains valid. However, instead of  Eq.~(\ref{eq21}), for
the plasma model we have
\begin{equation}
\lim_{\xi\to 0}\ve^{(\pm 1)}(i\xi)\xi^2=\omega_{p,\pm 1}^2,
\label{eq25}
\end{equation}
\noindent
and this leads to a nonzero contribution of the TE mode.
{}From Eqs.~(\ref{eq3})--(\ref{eq5}) one obtains
\begin{eqnarray}
&&
r_{\rm TE}^{(0,\pm 1)}(0,k_{\bot})=
-r_{\rm TE}^{(\pm 1,\pm 2)}(0,k_{\bot})
\nonumber \\
&&
\phantom{r_{\rm TE}^{(0,\pm 1)}(0,k_{\bot})}=
\frac{k_{\bot}-
(k_{\bot}^2+\omega_{p,\pm 1}^2/c^2)^{1/2}}{k_{\bot}+
(k_{\bot}^2+\omega_{p,\pm 1}^2/c^2)^{1/2}}
\equiv
r_{{\rm TE},p}^{(\pm 1)}(k_{\bot}),
\label{eq26} \\
&&
R_{\rm TE}^{(\pm )}(0,k_{\bot})=
\frac{r_{{\rm TE},p}^{(\pm 1)}(k_{\bot})\left\{1-
\exp[-2(k_{\bot}^2+\omega_{p,\pm 1}^2/c^2)^{1/2}d_{\pm}]\right\}}{1-
{r_{{\rm TE},p}^{(\pm 1)}}^2(k_{\bot})
\exp[-2(k_{\bot}^2+\omega_{p,\pm 1}^2/c^2)^{1/2}d_{\pm}]}.
\nonumber
\end{eqnarray}
\noindent
It is convenient to express the last reflection coefficient
in terms of the variable $y$
\begin{equation}
R_{\rm TE}^{(\pm )}(0,y)=
\frac{r_{{\rm TE},p}^{(\pm 1)}(y)\left\{1-
\exp[-d_{\pm}(y^2+{\tilde\omega_{p,\pm 1}}^2)^{1/2}/a]\right\}}{1-
{r_{{\rm TE},p}^{(\pm 1)}}^2(y)
\exp[-d_{\pm}(y^2+{\tilde\omega_{p,\pm 1}}^2)^{1/2}/a]},
\label{eq27}
\end{equation}
\noindent
where $\tilde\omega_{p,\pm 1}=2a\omega_{p,\pm 1}/c$ and
\begin{equation}
r_{{\rm TE},p}^{(\pm 1)}(y)=\frac{y-
\sqrt{y^2+{\tilde\omega_{p,\pm 1}}^2}}{y+
\sqrt{y^2+{\tilde\omega_{p,\pm 1}}^2}}.
\label{eq28}
\end{equation}
\noindent
Furthermore, we introduce the penetration depth of the
electromagnetic oscillations into metals of films
$\delta_{\pm 1}=c/\omega_{p,\pm 1}$ (it is equal to
approximately 22\,nm for Au). It is assumed that
$\delta_{\pm 1}\ll a$, so that the parameter
$\beta_{\pm 1}=\delta_{\pm 1}/(2a)\ll 1$. Expressing
Eq.~(\ref{eq27}) in terms of the quantities $\beta_{\pm 1}$
and $\delta_{\pm 1}$ and expanding in powers of $\beta_{\pm 1}$,
we arrive at
\begin{equation}
R_{\rm TE}^{(\pm )}(0,y)\approx
\frac{(-1+2\beta_{\pm 1}y)[1-
\exp(-2d_{\pm}/\delta_{\pm 1})]}{1-\exp(-2d_{\pm}/\delta_{\pm 1})
+4\beta_{\pm 1}y\exp(-2d_{\pm}/\delta_{\pm 1})}.
\label{eq29}
\end{equation}

The following calculations depend on the values of our parameters.
For sufficiently thick films satisfying the condition
\begin{equation}
\frac{d_{\pm}}{\delta_{\pm 1}}\gg 2\beta_{\pm 1}=
\frac{\delta_{\pm 1}}{a}
\label{eq30}
\end{equation}
\noindent
one can expand the right-hand side of Eq.~(\ref{eq29}) in powers
of a small parameter $\delta_{\pm 1}/a$ and obtain
\begin{equation}
R_{\rm TE}^{(\pm )}(0,y)\approx
-1+2\beta_{\pm 1}y\coth\frac{d_{\pm}}{\delta_{\pm 1}}.
\label{eq31}
\end{equation}
\noindent
Note that Eq.~(\ref{eq30}) is satisfied also for films of any
thickness if the separation distance $a$ is sufficiently large.
By contrast, for very thin films satisfying the condition
\begin{equation}
\frac{d_{\pm}}{\delta_{\pm 1}}\ll
\frac{\delta_{\pm 1}}{a}\ll 1,
\label{eq32}
\end{equation}
\noindent
one can expand the exponents in Eq.~(\ref{eq29}) in powers
of $d_{\pm}/\delta_{\pm 1}$.
This leads to the result
\begin{equation}
R_{\rm TE}^{(\pm )}(0,y)\approx-
\frac{d_{\pm}/\delta_{\pm 1}}{d_{\pm}/\delta_{\pm 1}
+2\beta_{\pm 1}y}.
\label{eq33}
\end{equation}

If both films satisfy Eq.~(\ref{eq30}) one finds from
Eq.~(\ref{eq31})
\begin{equation}
R_{\rm TE}^{(+)}(0,y)R_{\rm TE}^{(-)}(0,y)\approx
1-2\beta_{1}y\coth\frac{d_{+}}{\delta_{1}}
-2\beta_{-1}y\coth\frac{d_{-}}{\delta_{-1}}.
\label{eq34}
\end{equation}
\noindent
Substituting Eq.~(\ref{eq34}) in Eq.~(\ref{eq2}) written in
terms of the variable $y$ and performing integration in $y$, we
obtain for the TE contribution to the Casimir free energy
\begin{equation}
{\cal F}_{\rm TE}(a,T)\approx -\frac{k_BT\zeta(3)}{16\pi a^2}
\left[1-2\left(\frac{\delta_{1}}{a}\coth\frac{d_{+}}{\delta_{1}}
+\frac{\delta_{-1}}{a}\coth\frac{d_{-}}{\delta_{-1}}\right)\right].
\label{eq35}
\end{equation}
\noindent
The contribution of the TM mode to the free energy
 is given by
Eq.~(\ref{eq24}). Then the total Casimir free energy
per unit area in the
classical limit is given by
\begin{equation}
{\cal F}_{p}(a,T)\approx -\frac{k_BT\zeta(3)}{8\pi a^2}
\left[1-\frac{\delta_{1}}{a}\coth\frac{d_{+}}{\delta_{1}}
-\frac{\delta_{-1}}{a}\coth\frac{d_{-}}{\delta_{-1}}\right].
\label{eq36}
\end{equation}
\noindent
Here the magnitude of the main contribution is twice that
in Eq.~(\ref{eq24}) and our result depends on
the film properties.

Let now the lower film satisfies Eq.~(\ref{eq30}) and the
upper film satisfies Eq.~(\ref{eq32}). In this case
using Eqs.~(\ref{eq31}) and (\ref{eq33}) in the leading order
we obtain
\begin{equation}
R_{\rm TE}^{(+)}(0,y)R_{\rm TE}^{(-)}(0,y)\approx-
\frac{d_{+}/\delta_{1}}{d_{+}/\delta_{1}
+2\beta_{1}y}.
\label{eq37}
\end{equation}
\noindent
Then, expanding the logarithm under the integral in
Eq.~(\ref{eq2})
and integrating using Eq.~(\ref{eq37}) one arrives at
the result
\begin{equation}
{\cal F}_{\rm TE}(a,T)\approx -\frac{k_BTd_{+}}{16\pi a\delta_{1}^2}
\left[1-\frac{d_{+}a}{\delta_{1}^2}e^{d_{+}a/\delta_{1}^2}
\Gamma\left(0,\frac{d_{+}a}{\delta_{1}^2}\right)\right],
\label{eq38}
\end{equation}
\noindent
where $\Gamma(x,y)$ is the incomplete gamma function. Note that
due to Eq.~(\ref{eq32}) the correction to unity on the right-hand
side of  Eq.~(\ref{eq38}) is negligibly small. By adding the
contribution of the TM mode, one arrives to the total Casimir
free energy per unit area
\begin{equation}
{\cal F}_{p}(a,T)\approx -\frac{k_BT\zeta(3)}{16\pi a^2}
\left[1+\frac{d_{+}a}{\zeta(3)\delta_{1}^2}\right].
\label{eq39}
\end{equation}

Finally we consider the two films satisfying Eq.~(\ref{eq32}).
In this case from Eq.~(\ref{eq33}) we have
\begin{equation}
R_{\rm TE}^{(+)}(0,y)R_{\rm TE}^{(-)}(0,y)\approx
\frac{d_{+}d_{-}}{\delta_{1}^2\delta_{-1}^2}\,
\frac{a^2}{(d_{+}a/\delta_{1}^2+y)(d_{-}a/\delta_{-1}^2+y)}.
\label{eq40}
\end{equation}
\noindent
After integration according to Eq.~(\ref{eq2}), one
obtains
\begin{eqnarray}
&&
{\cal F}_{\rm TE}(a,T)\approx -\frac{k_BT}{16\pi}\,
\frac{d_{+}d_{-}}{d_{-}\delta_{1}^2-d_{+}\delta_{-1}^2}
\nonumber \\
&&
~~\times
\left[\frac{d_{-}}{\delta_{-1}^2}e^{d_{-}a/\delta_{-1}^2}
\Gamma\left(0,\frac{d_{-}a}{\delta_{-1}^2}\right)-
\frac{d_{+}}{\delta_{1}^2}e^{d_{+}a/\delta_{1}^2}
\Gamma\left(0,\frac{d_{+}a}{\delta_{1}^2}\right)\right].
\label{eq41}
\end{eqnarray}
\noindent
The total free energy is obtained by adding Eq.~(\ref{eq24})
to this equation. It is seen that the contribution of the TE
mode is much smaller than the contribution of the TM mode
given by Eq.~(\ref{eq24}). The respective results for the
Casimir pressure are obtained from Eqs.~(\ref{eq24}),
(\ref{eq36}), (\ref{eq39}) and (\ref{eq41}) by the negative
differentiation with respect to $a$.

Note that the free energy ${\cal F}_p$ in all cases considered
depends on film thicknesses, as opposed to ${\cal F}_D$
obtained using the Drude model. This is because the classical
limit for the Drude metals does not depend on any metallic
property due to $r_{\rm TM}(0,k_{\bot})=1$ and
$r_{\rm TE}(0,k_{\bot})=0$.
As to the plasma model, $r_{\rm TE}(0,k_{\bot})\neq 0$ and
depends on the penetration depth of electromagnetic oscillations
into the metal. Then it is not surprising that, under different
relationships among the penetration depths and film thicknesses,
different classical limits considered above are possible which
depend on both the penetration depths and film thicknesses.

\section{Films deposited on substrates}

In this section we consider the classical Casimir interaction of
a thin film deposited on a substrate with thick plates
(semispaces) made of different materials or with another thin
film also deposited on a substrate.
We begin from the case of a dielectric film having the
dielectric permittivity $\ve^{(1)}(\omega)$ deposited on a
dielectric semispace having the dielectric permittivity
$\ve^{(2)}(\omega)$ interacting with a dielectric semispace having the
dielectric permittivity $\ve^{(-1)}(\omega)=\ve^{(-2)}(\omega)$
(see Fig.~\ref{fg1}). Then from Eqs.~(\ref{eq3})--(\ref{eq5})
we obtain
\begin{eqnarray}
&&
R_{\rm TE}^{(\pm)}(0,k_{\bot})=0,\quad
r_{\rm TM}^{(1,2)}(0,k_{\bot})=
\frac{\ve_0^{(2)}-\ve_0^{(1)}}{\ve_0^{(2)}+\ve_0^{(1)}}
\equiv r_0^{(2,1)},
\nonumber \\
&&
r_{\rm TM}^{(-1,-2)}(0,k_{\bot})=0,\quad
r_{\rm TM}^{(0,\pm 1)}(0,k_{\bot})=r_0^{(\pm 1)},
\label{eq42} \\
&&
R_{\rm TM}^{(+)}(0,k_{\bot})=\frac{r_0^{(1)}+
r_0^{(2,1)}e^{-2k_{\bot}d_{+}}}{1+r_0^{(1)}
r_0^{(2,1)}e^{-2k_{\bot}d_{+}}}
\quad
R_{\rm TM}^{(-)}(0,k_{\bot})=r_0^{(-1)}.
\nonumber
\end{eqnarray}

Substituting Eq.~(\ref{eq42}) in  Eq.~(\ref{eq2}), using the
variable $y$ and expanding in powers of a small parameter
$d_{+}/a$ under the integral one finds
\begin{eqnarray}
&&
{\cal F}(a,T)\approx\frac{k_BT}{16\pi a^2}\int_{0}^{\infty}
\!\!\!ydy\ln\left\{
\vphantom{\left[\frac{{\ve_0^{(2)}}^2-{\ve_0^{(1)}}^2}{\ve_0^{(1)}} \right]}
1-r_0^{(-1)}\right.
\nonumber \\
&&
~~~~~~~~\left.
\times\left[
r_0^{(2)}
-\frac{d_{+}}{a}\,
\frac{{\ve_0^{(2)}}^2-{\ve_0^{(1)}}^2}{\ve_0^{(1)}(1+
\ve_0^{(2)})^2}\,y\right]e^{-y}\right\},
\label{eq43}
\end{eqnarray}
\noindent
where $r_0^{(2)}$ is defined in the same way as $r_0^{(-2)}$
in Eq.~(\ref{eq13}).
Integration in Eq.~(\ref{eq43}) leads to
\begin{equation}
{\cal F}(a,T)\approx\frac{k_BT{\rm Li}_3(r_0^{(-1)}r_0^{(2)})}{16\pi a^2}
\left[1-2\frac{{\ve_0^{(2)}}^2-{\ve_0^{(1)}}^2}{\ve_0^{(1)}
(\ve_0^{(2)}-1)}\,\frac{d_{+}}{a}\right],
\label{eq44}
\end{equation}
\noindent
where ${\rm Li}_n(z)$ is the polylogarithm function.

In a similar way one can find the Casimir free energy for a
dielectric film deposited on thick dielectric plate and
interacting with a metallic semispace. Here the contribution
of the TE mode is again equal to zero due to
$R_{\rm TE}^{(+)}(0,k_{\bot})=0$ and
$R_{\rm TM}^{(-)}(0,k_{\bot})=1$. As a result,
\begin{equation}
{\cal F}(a,T)\approx
-\frac{k_BT{\rm Li}_3(r_0^{(2)})}{16\pi a^2}
\left[1-2\frac{{\ve_0^{(2)}}^2-{\ve_0^{(1)}}^2}{\ve_0^{(1)}
(\ve_0^{(2)}-1)}\,\frac{d_{+}}{a}\right]
\label{eq45}
\end{equation}
\noindent
independently of the model of a metal used.

The next case to consider is the dielectric film deposited
on a metallic plate and interacting with a dielectric plate.
The TE mode, again, does not contribute, this time due to
$R_{\rm TE}^{(-)}(0,k_{\bot})=0$, and the result does not depend
on a model of metal. From Eqs.~(\ref{eq3})--(\ref{eq5}) one
finds
\begin{equation}
R_{\rm TM}^{(+)}(0,k_{\bot})=\frac{r_0^{(1)}+
e^{-2k_{\bot}d_{+}}}{1+r_0^{(1)}e^{-2k_{\bot}d_{+}}},
\quad
R_{\rm TM}^{(-)}(0,k_{\bot})=r_0^{(-1)}.
\label{eq46}
\end{equation}
\noindent
Substituting this in Eq.~(\ref{eq2}),  using the
variable $y$ and expanding in powers of
$d_{+}/a$, we obtain
\begin{eqnarray}
&&
{\cal F}(a,T)\approx\frac{k_BT}{16\pi a^2}\int_{0}^{\infty}
\!\!\!ydy
\label{eq47} \\
&&
~~~\times
\left[\ln(1-r_0^{(-1)}e^{-y})+
\frac{r_0^{(-1)}}{\ve_0^{(1)}}\,\frac{d_{+}}{a}\,
\frac{y}{e^y-r_0^{(-1)}}\right].
\nonumber
\end{eqnarray}
\noindent
After the integration of Eq.~(\ref{eq47}) one arrives at
\begin{equation}
{\cal F}(a,T)\approx
-\frac{k_BT{\rm Li}_3(r_0^{(-1)})}{16\pi a^2}
\left[1-\frac{2}{\ve_0^{(1)}}
\,\frac{d_{+}}{a}\right].
\label{eq48}
\end{equation}

The case of a dielectric film deposited on a metallic plate and
interacting with a metallic plate is a bit more complicated.
Here, the result depends on the model of metal.
If the low-frequency behavior of $\ve^{(2)}(\omega)$ and
$\ve^{(-1)}(\omega)=\ve^{(-2)}(\omega)$ is described by the
Drude model, we have $R_{\rm TE}^{(-)}(0,k_{\bot})=0$ and
$r_0^{(-1)}=1$. Then, repeating all calculations, as in the
previous case, one finds
\begin{equation}
{\cal F}_D(a,T)\approx
-\frac{k_BT\zeta(3)}{16\pi a^2}
\left[1-\frac{2}{\ve_0^{(1)}}
\,\frac{d_{+}}{a}\right].
\label{eq49}
\end{equation}
\noindent
Taking into account that ${\rm Li}_3(1)=\zeta(3)$, this is in agreement
with Eq.~(\ref{eq48}).

Now we admit that the low-frequency behavior of $\ve^{(2)}(\omega)$ and
$\ve^{(-1)}(\omega)=\ve^{(-2)}(\omega)$ is described by the
plasma model. In this case the contribution of the TM mode
remains unchanged, i.e., it is given by Eq.~(\ref{eq49}).
In addition, the contribution of the TE mode becomes nonzero
\begin{eqnarray}
&&
R_{\rm TE}^{(+)}(0,k_{\bot})=
\frac{k_{\bot}-
(k_{\bot}^2+\omega_{p,2}^2/c^2)^{1/2}}{k_{\bot}+
(k_{\bot}^2+\omega_{p,2}^2/c^2)^{1/2}}\,
e^{-2k_{\bot}d_{+}},
\nonumber \\
&&
R_{\rm TE}^{(-)}(0,k_{\bot})=
r_{{\rm TE},p}^{(-1)}(k_{\bot}),
\label{eq50}
\end{eqnarray}
\noindent
where $r_{{\rm TE},p}^{(-1)}(k_{\bot})$ is defined in
Eq.~(\ref{eq26})
and $\omega_{p,2}$ is the plasma frequency of the substrate
metal. Using the variable $y$ and performing the expansion
in powers of $d_{+}/a$, we obtain
\begin{eqnarray}
&&
R_{\rm TE}^{(+)}(0,k_{\bot})\approx
-1+2\beta_2y+\frac{d_{+}}{a}y,
\nonumber \\
&&
R_{\rm TE}^{(-)}(0,k_{\bot})\approx
-1+2\beta_{-1}y,
\label{eq51}
\end{eqnarray}
\noindent
where $\beta_2\equiv\delta_2/(2a)$, $\delta_2=c/\omega_{p,2}$
and $\beta_{-1}$ is defined in Sec.~III.
Using Eqs.~(\ref{eq2}) and (\ref{eq51}), to the first order of
small parameters $\beta_{-1}$, $\beta_2$ and $d_{+}/a$, for
the contribution of the TE mode to the Casimir free energy
we have
\begin{eqnarray}
&&
{\cal F_{\rm TE}}(a,T)\approx\frac{k_BT}{16\pi a^2}\int_{0}^{\infty}
\!\!\!ydy
\label{eq52} \\
&&
~~~\times
\ln\left\{1-\left[
1-2(\beta_2+\beta_{-1})y
-\frac{d_{+}}{a}y
\right]e^{-y}\right\}.
\nonumber
\end{eqnarray}
\noindent
Expanding the logarithm in powers of the same small parameters
and integrating, we obtain
\begin{equation}
{\cal F}_{\rm TE}(a,T)\approx\frac{k_BT\zeta(3)}{16\pi a^2}\left[
-1+4(\beta_2+\beta_{-1})
+2\frac{d_{+}}{a}
\right].
\label{eq53}
\end{equation}
\noindent
By adding to this equation the contribution of the TM mode in
Eq.~(\ref{eq49}), the final result is
\begin{equation}
{\cal F}_p(a,T)\approx-\frac{k_BT\zeta(3)}{8\pi a^2}\left[
1-\frac{\delta_2+\delta_{-1}}{a}-
\frac{\ve_0^{(1)}+1}{\ve_0^{(1)}}\,\frac{d_{+}}{a}
\right].
\label{eq54}
\end{equation}
\noindent
Thus, in the case of the plasma metals, the magnitude of the
main contribution is larger by a factor of 2, as compared to
Eq.~(\ref{eq49}), and the correction term depends on the
penetration depth of electromagnetic oscillations into
a metal.

Now we consider the classical Casimir interaction of two
thin films each of which is deposited on a thick plate
(semispace). We begin with two dielectric films with
dielectric permittivities $\ve^{(\pm 1)}$ deposited on
the dielectric semispaces having the
permittivities $\ve^{(\pm 2)}$, respectively.
Calculations for this case are presented in Appendix A.
Substituting Eq.~(\ref{eq57}) in
 Eq.~(\ref{eq2}), expanding the logarithm and performing
the integration, one arrives at
\begin{eqnarray}
&&
{\cal F}(a,T)\approx
-\frac{k_BT{\rm Li}_3(r_0^{(2)}r_0^{(-2)})}{16\pi a^2}
\label{eq58} \\
&&
~~~~~~\times\left\{1-2
\frac{{\ve_0^{(2)}}^2-{\ve_0^{(1)}}^2}{\ve_0^{(1)}
[{\ve_0^{(2)}}^2-1]}\,\frac{d_{+}}{a}-2
\frac{{\ve_0^{(-2)}}^2-{\ve_0^{(-1)}}^2}{\ve_0^{(-1)}
[{\ve_0^{(-2)}}^2-1]}\,\frac{d_{-}}{a}\right\}.
\nonumber
\end{eqnarray}

If we now replace the lower dielectric plate (semispace) with
a metallic one, the free energy of the Casimir interaction
between two dielectric films does not depend on the model of
a metal used. Similar to the respective case in Sec.~III,
the result can be obtained from Eq.~(\ref{eq58}) in the
limiting case ${\ve_0^{(-2)}}^2\to\infty$:
\begin{eqnarray}
&&
{\cal F}(a,T)\approx
-\frac{k_BT{\rm Li}_3(r_0^{(2)})}{16\pi a^2}
\label{eq59} \\
&&
~~~~~~\times\left\{1-2
\frac{{\ve_0^{(2)}}^2-{\ve_0^{(1)}}^2}{\ve_0^{(1)}
[{\ve_0^{(2)}}^2-1]}\,\frac{d_{+}}{a}-2
\frac{2}{\ve_0^{(-1)}}
\,\frac{d_{-}}{a}\right\}.
\nonumber
\end{eqnarray}

The case when both dielectric films are deposited on
the metallic semispaces is a bit more cumbersome. Here the
result depends on the model of a metal.
If the Drude model (\ref{eq19}) is used, one has
\begin{equation}
R_{\rm TE}^{(\pm)}(0,y)=0,\quad
R_{\rm TM}^{(\pm)}(0,y)\approx 1-
\frac{1}{\ve_0^{(\pm 1)}}\,\frac{d_{\pm}}{a}y.
\label{eq60}
\end{equation}
\noindent
{}From this it follows:
\begin{equation}
R_{\rm TM}^{(+)}(0,y)R_{\rm TM}^{(-)}(0,y)
\approx 1-
\frac{1}{\ve_0^{(1)}}\,\frac{d_{+}}{a}y-
\frac{1}{\ve_0^{(-1)}}\,\frac{d_{-}}{a}y
\label{eq61}
\end{equation}
\noindent
and from Eq.~(\ref{eq2})
\begin{equation}
{\cal F}_D(a,T)\approx -\frac{k_BT\zeta(3)}{16\pi a^2}\left[1-
\frac{2}{\ve_0^{(1)}}\,\frac{d_{+}}{a}-
\frac{2}{\ve_0^{(-1)}}\,\frac{d_{-}}{a}\right].
\label{eq62}
\end{equation}
\noindent
If the metal of semispaces is described by the plasma model
(\ref{eq20}), the reflection coefficients
$R_{\rm TE}^{(\pm)}(0,y)$
become nonzero.
Calculation details for this case are given in Appendix A.
By adding the contribution of the TM mode [which is the same
as in Eq.~(\ref{eq62})] to the contribution of the TE mode
presented in Eq.~(\ref{eq65}), we obtain
\begin{equation}
{\cal F}_p(a,T)\approx -\frac{k_BT\zeta(3)}{8\pi a^2}\left(1-
\frac{\delta_2+\delta_{-2}}{a}
-\frac{\ve_0^{(1)}+1}{\ve_0^{(1)}}\,\frac{d_{+}}{a}
-\frac{\ve_0^{(-1)}+1}{\ve_0^{(-1)}}\,\frac{d_{-}}{a}
\right).
\label{eq66}
\end{equation}

The Casimir pressures for all cases considered in this section
can be easily obtained from Eqs.~(\ref{eq44}), (\ref{eq45}),
(\ref{eq48}), (\ref{eq49}),
(\ref{eq54})--(\ref{eq59}),
(\ref{eq62}),  and (\ref{eq66}) by the differentiation
with respect to $a$. Note also that the case of two metallic
films covering semispaces made of any material is equivalent to
two metallic semispaces if the films are sufficiently thick.
Because of this we do not consider it here.

\section{Applications to the Casimir interaction of two
graphene sheets}

As mentioned in Sec.~II, for two films described by the
dielectric permittivity the classical regime holds for
separations larger than a few micrometers. For two sheets
of pristine graphene
 described by the polarization tensor in the framework
of Dirac model the classical regime starts at much shorter
separations.\cite{22} In this regime the free energy of the
Casimir interaction per unit area of graphene sheet is
again given by Eq.~(\ref{eq2}) where the reflection
coefficients
$R_{\alpha}^{(+)}(0,k_{\bot})=R_{\alpha}^{(-)}(0,k_{\bot})
\equiv R_{\alpha}(0,k_{\bot})$ take the form\cite{23,31}
\begin{eqnarray}
&&
R_{\rm TM}(0,k_{\bot})=
\frac{\Pi_{00}(0,k_{\bot})}{\Pi_{00}(0,k_{\bot})+
2\hbar k_{\bot}},
\label{eq67} \\
&&
R_{\rm TE}(0,k_{\bot})=
-\frac{\Pi_{\rm tr}(0,k_{\bot})-
\Pi_{00}(0,k_{\bot})}{\Pi_{\rm tr}(0,k_{\bot})-
\Pi_{00}(0,k_{\bot})+2\hbar k_{\bot}}.
\nonumber
\end{eqnarray}
\noindent
Here, $\Pi_{00}$ is the 00-component of the polarization
tensor in (2+1)-dimensional space-time and trace stands
for the sum of spatial components ${\Pi_1}^1$ and
${\Pi_2}^2$ all taken at zero frequency.
The explicit expressions are:\cite{23,30,31,53}
\begin{eqnarray}
&&
\Pi_{00}(0,k_{\bot})=\frac{16\alpha k_BTc}{v_F^2}
\int_{0}^{1}\!\!dx\ln[2\cosh\varphi(k_{\bot},x)],
\nonumber \\
&&
\Pi_{\rm tr}(0,k_{\bot})=\Pi_{00}(0,k_{\bot})+
\frac{8\alpha\hbar v_Fk_{\bot}}{c}
\int_{0}^{1}\!\!dx\sqrt{x(1-x)}\tanh\varphi(k_{\bot},x),
\label{eq67a}
\end{eqnarray}
\noindent
where
\begin{equation}
\varphi(k_{\bot},x)=\frac{\hbar v_Fk_{\bot}}{2k_BT}
\sqrt{x(1-x)},
\label{eq67b}
\end{equation}
\noindent
$\alpha=e^2/(\hbar c)$ is the fine structure constant and
$v_F\approx c/300$ is the Fermi velocity.

Substituting Eq.~(\ref{eq67}) in Eq.~(\ref{eq2}), one
arrives to the Casimir free energy per unit area and
pressure between two sheets of gapless graphene in the
classical limit\cite{22,23,31}
\begin{eqnarray}
&&
{\cal F}(a,T)\approx -\frac{k_BT\zeta(3)}{16\pi a^2}
\left[1-\frac{1}{4\alpha\ln 2}
\left(\frac{v_F}{c}\right)^2
\frac{\hbar c}{ak_BT}\right],
\nonumber \\
&&
P(a,T)\approx -\frac{k_BT\zeta(3)}{8\pi a^3}
\left[1-\frac{3}{8\alpha\ln 2}
\left(\frac{v_F}{c}\right)^2
\frac{\hbar c}{ak_BT}\right].
\label{eq68}
\end{eqnarray}
\noindent
 Numerical computations
show that the asymptotic expressions (\ref{eq68}) are
applicable at much smaller separations than the asymptotic
expressions obtained in Secs.~II--IV for usual dielectric and
metallic films. Thus, at $T=300\,$K the pressure values
calculated using Eq.~(\ref{eq68}) agree in the limits of 1\% and
5\% with the total pressure computed by taking into account the
contributions from all Matsubara frequencies at separations
$a>370\,$nm and $a>150\,$nm, respectively.
Note that the dependence of the free energy on separation
proportional to $a^{-2}$ was also obtained in
Refs.~\cite{22,28} and \cite{53a}.

The main, first, term on the right-hand side of Eq.~(\ref{eq68})
does not contain the Planck constant and, thus, corresponds to
the classical limit.\cite{37} We emphasize that for two
dielectric films with no free charge carriers the classical
limit for the free energy in Eqs.~(\ref{eq10}) and (\ref{eq11})
contains the thicknesses of the films and is inversely proportional to
the fourth power of separation. For this reason, if graphene
sheets are modeled as two insulator films (see, e.g.,
Refs.~\cite{27,54}), the computed free energy deviates
significantly from that computed using the polarization tensor
within a wide region of separations. It was
hypothesized\cite{27,54} that the account of thermally
excited charge carriers at nonzero temperature (which are
always present in dielectric material in the form of
dc conductivity) should change the long-distance power
$n=4$ as in Eqs.~(\ref{eq10}) and (\ref{eq11}), to $n=2$,
as in Eq.~(\ref{eq68}).

In fact the dc conductivity of dielectrics can be
described by the Drude-type additional term in the
dielectric permittivity which leads to Eq.~(\ref{eq21})
and then to Eqs.~(\ref{eq22}) and (\ref{eq23}).
As a result, the free energy of two dielectric films with
account of dc conductivity, as well as of two metallic
films described by the Drude model, is given by
Eq.~(\ref{eq24}). This is in agreement with the main,
first, term on the right-hand side of Eq.~(\ref{eq68})
and does not depend on the film thicknesses. Thus,
the description of a gapless graphene at nonzero
temperature as a dielectric film possessing some
dc conductivity is really a better model leading
to a correct classical limit. It should be noted,
however, that for dielectric films with account of
dc conductivity the asymptotic regime (\ref{eq24})
is achieved starting from separations of a few
micrometers, whereas for graphene at more than an
order of magnitude shorter separations.
It is worth mentioning also that for two dielectric
plates with account of dc conductivity the Lifshitz theory
violates\cite{2,6,55} the third law of thermodynamics
(the Nernst heat theorem).

For graphene modeled as two
metallic films described by the plasma model under the
condition (\ref{eq30}), the Casimir free energy is given
by Eq.~(\ref{eq36}). Here the magnitude of the main term
is larger by a factor of two than for graphene sheets
in Eq.~(\ref{eq68}). If, however, at least one film is
described by the plasma model under the condition (\ref{eq32}),
we again arrive to the same main term as in Eq.~(\ref{eq68})
obtained for graphene sheets [see, e.g., Eq.~(\ref{eq39})].
Thus, both the Drude and the plasma models can be used to obtain
the classical limit for graphene. Furthermore, the Drude-type
dielectric permittivity leads to exactly zero contribution
of the TE mode at zero frequency, whereas for graphene it is,
although small, but not equal to zero. This qualitatively
likens graphene to a metallic film described by the plasma
model. By and large both the Drude and the plasma dielectric
permittivities do not provide a correct quantitative
description of such a unique two-dimensional system as
graphene.

Now we compare the results of Sec.~II for the Casimir
interaction between a thin film and a thick plate (semispace)
with respective results for a graphene sheet.
Using the formalism of a polarization tensor, the Casimir
free energy of graphene interacting with a dielectric
semispace was calculated in Ref.~\cite{30}.
In the classical limit for a gapless graphene it is
given by the asymptotic expression\cite{23,30}
\begin{equation}
{\cal F}(a,T)\approx -
\frac{k_BT{\rm Li}_3(r_0^{(-2)})}{16\pi a^2}
\left[1-\frac{1}{8\alpha\ln 2}
\left(\frac{v_F}{c}\right)^2
\frac{\hbar c}{ak_BT}\right].
\label{eq69}
\end{equation}
\noindent
This demonstrates another dependence on separation than in the
first equation (\ref{eq15}) obtained for a dielectric film of
thickness $d_{+}$ interacting with a dielectric semispace.

The next step is to compare Eq.~(\ref{eq69}) with the free
energy of the classical
Casimir interaction between a metallic film and a
dielectric semispace. The latter quantity is the same as
for metallic and dielectric semispaces. It does not depend
on the model of metal used and is given by\cite{56}
\begin{equation}
{\cal F}(a,T)\approx -
\frac{k_BT{\rm Li}_3(r_0^{(-2)})}{16\pi a^2}
\label{eq70}
\end{equation}
\noindent
in accordance with the main contribution to Eq.~(\ref{eq69})
obtained for a graphene--semispace interaction.

The classical Casimir interaction between a graphene sheet
and a metallic semispace is of interest because of possible
dependence on the model of metal. If the semispace is described
by the Drude model, the free energy of graphene--semispace
interaction is given by\cite{22,23,30}
\begin{equation}
{\cal F}(a,T)\approx -
\frac{k_BT\zeta(3)}{16\pi a^2}
\left[1-\frac{1}{8\alpha\ln 2}
\left(\frac{v_F}{c}\right)^2
\frac{\hbar c}{ak_BT}\right].
\label{eq71}
\end{equation}
\noindent
If the semispace is described by the plasma model, only a
negligible small correction to Eq.~(\ref{eq71}) is
added.\cite{30}

Let us compare this result with the classical Casimir
interaction of the material film with a metallic
semispace, both describe by some dielectric permittivities.
If the film is made of insulator material, the classical
Casimir free energy of its interaction with a metallic plate
is given by the first equation (\ref{eq18}) and does not
depend on the model of metal. As can be seen in
Eq.~(\ref{eq18}), it depends on the film thickness $d_{+}$
and differs essentially from the main contribution to
Eq.~(\ref{eq71}). Alternatively, if the film is metallic the
classical Casimir free energy depends on the model of a metal.
If both the film and the semispace metals are described at low
frequencies by the Drude model, the result is\cite{2,6}
\begin{equation}
{\cal F}_D(a,T)\approx -
\frac{k_BT\zeta(3)}{16\pi a^2},
\label{eq72}
\end{equation}
\noindent
i.e., the same as the main contribution in Eq.~(\ref{eq71}).
One should remember, however, that the Lifshitz theory
combined with the Drude model violates the Nernst heat
theorem.\cite{43,44,45}
If both metals are described at low
frequencies by the plasma model under the condition (\ref{eq30}),
one obtains ${\cal F}_p\approx 2{\cal F}_D$
in disagreement with
the main contribution to Eq.~(\ref{eq71}). When the
condition (\ref{eq32}) is satisfied,
one obtains ${\cal F}_p\approx {\cal F}_D$
in agreement with Eq.~(\ref{eq71}) found using the
description of graphene by means of the polarization tensor.

\section{Conclusions and discussion}

In the foregoing we have considered the classical Casimir
interaction between two thin material films made of dielectrics
and metals, between a film and a thick plate (semispace),
and between two films deposited on substrates.
The consideration of a classical limit, where the interaction
does not depend on the Planck constant, presents some
advantages, because all results can be found in simple
analytic form. The problem of the Casimir interaction
between thin films has become topical because of
increasing interest to the dispersion interactions of
graphene. In this area many results were obtained by using
the Lifshitz theory and describing graphene sheets in
close analogy with material films of nonzero thickness
possessing some model dielectric permittivity.
Alternatively (and more fundamentally) the reflection
coefficients of the electromagnetic oscillations on
graphene described by the Dirac model were expressed
via the polarization tensor with no recourse to the
concept of dielectric permittivity. Some of the results
obtained using both these approaches were found
to be in mutual disagreement.

In this paper we have shown that the classical limit of
the Casimir interaction for two dielectric films with no
free charge carriers is different from the classical limit
for two dielectric semispaces and depends on film thicknesses.
The respective Casimir free energy decreases as an inversely
fourth power of separation instead of the inversely second
power, as it holds in the classical limit
for two dielectric semispaces and for two
graphene sheets described by the polarization tensor in the
framework of the Dirac model. Thus, modeling of the pristine
graphene as an insulating film should be considered as
unrealistic.

Another conclusion is that one obtains the correct main contribution
to the classical Casimir free energy when the graphene sheets are
modeled as material films with metallic-type dielectric
permittivity. This, however, is achieved at separations of a few
micrometers, i.e., at much larger separations than the classical
regime is achieved\cite{22} for graphene sheets described by the Dirac
model. As a result, at moderate separations between graphene
sheets, which are of most experimental interest, the theoretical
predictions using the model of graphene as a dielectric
possessing  some dc conductivity cannot be considered as
enough reliable.

One can also conclude  that it is most prospective to investigate
the dispersion interactions between graphene sheets, both
isolated and deposited on substrates, using the formalism of
the polarization tensor. In this connection it is desirable to
generalize the quantum electrodynamical formalism to the case
when the two-dimensional layer described by the polarization
tensor
is sandwiched between two material plates described by two
different dielectric permittivities.

\section*{APPENDIX A}
\renewcommand{\theequation}{A\arabic{equation}}
\setcounter{equation}{0}

Here we present a few details of analytic derivations related
to Sec.~IV. We begin from the case of two dielectric films each
of which is deposited on a dielectric semispace.
{}From Eqs.~(\ref{eq3})--(\ref{eq5}) it follows
\begin{eqnarray}
&&
R_{\rm TM}^{(\pm)}(0,k_{\bot})=
\frac{r_0^{(\pm 1)}+r_0^{(\pm 2,\pm 1)}
e^{-2k_{\bot}d_{\pm}}}{1+r_0^{(\pm 1)}r_0^{(\pm 2,\pm 1)}
e^{-2k_{\bot}d_{\pm}}},
\nonumber \\
&&
R_{\rm TE}^{(\pm)}(0,k_{\bot})=0,
\label{eq55}
\end{eqnarray}
\noindent
where $r_0^{(\pm 1)}$ is defined in Eq.~(\ref{eq6}) and
$r_0^{(-2,-1)}$ is defined in the same way as
$r_0^{(2,1)}$ in Eq.~(\ref{eq42}). Substituting these
definitions in Eq.~(\ref{eq55}), using the variable $y$
and expanding in powers of small parameters
$d_{\pm}/a$, one obtains
\begin{equation}
R_{\rm TM}^{(\pm)}(0,y)\approx r_0^{(\pm 2)}-
\frac{{\ve_0^{(\pm 2)}}^2-{\ve_0^{(\pm 1)}}^2}{\ve_0^{(\pm 1)}
[\ve_0^{(\pm 2)}+1]^2}\,\frac{d_{\pm}}{a}\,y.
\label{eq56}
\end{equation}
\noindent
 {}From Eq.~(\ref{eq56}) it is easily
obtainable
\begin{eqnarray}
&&
R_{\rm TM}^{(+)}(0,y)R_{\rm TM}^{(-)}(0,y)
\approx r_0^{(2)}r_0^{(-2)}
\label{eq57} \\
&&
~~~~~
-
r_0^{(2)}\frac{{\ve_0^{(-2)}}^2-{\ve_0^{(-1)}}^2}{\ve_0^{(-1)}
[\ve_0^{(-2)}+1]^2}\,\frac{d_{-}}{a}\,y
-
r_0^{(-2)}\frac{{\ve_0^{(2)}}^2-{\ve_0^{(1)}}^2}{\ve_0^{(1)}
[\ve_0^{(2)}+1]^2}\,\frac{d_{+}}{a}\,y.
\nonumber
\end{eqnarray}
\noindent
This is used in the main text to obtain Eq.~(\ref{eq58}).

Now we present some calculation details for the case of two
dielectric films deposited on metallic semispaces described
 by the plasma model.
{}From Eqs.~(\ref{eq3})--(\ref{eq5}) we have
\begin{eqnarray}
&&
R_{\rm TE}^{(\pm)}(0,y)\approx r_{\rm TE}^{(\pm 1,\pm2)}(0,y)
e^{-d_{\pm}y/a}
\label{eq63} \\
&&
~~~\approx
(-1+2\beta_{\pm 2}y)\left(1-\frac{d_{\pm}}{a}y\right)
\approx -1+\frac{\delta_{\pm 2}}{a}y+\frac{d_{\pm}}{a}y,
\nonumber
\end{eqnarray}
\noindent
where $\delta_{\pm 2}=c/\omega_{p,\pm 2}$, $\omega_{p,\pm 2}$
are the plasma frequencies of the upper and lower metallic semispaces
and $\beta_{\pm 2}=\delta_{\pm 2}/(2a)$.
{}From Eq.~(\ref{eq63}) one has
\begin{equation}
R_{\rm TE}^{(+)}(0,y)R_{\rm TE}^{(-)}(0,y)\approx
1-\frac{\delta_2+\delta_{-2}+d_{+}+d_{-}}{a}y,
\label{eq64}
\end{equation}
\noindent
which, with the help of Eq.~(\ref{eq2}), leads to
\begin{equation}
{\cal F}_{\rm TE}(a,T)\approx -\frac{k_BT\zeta(3)}{16\pi a^2}\left(1-
2\frac{\delta_2+\delta_{-2}+d_{+}+d_{-}}{a}\right).
\label{eq65}
\end{equation}
\noindent
This is used to obtain Eq.~(\ref{eq66}) in the main text.

\end{document}